\documentclass{appolb}
\usepackage{epsfig}

\newcommand {\pom} {I\!\!P}
\newcommand {\pomsub} {{\scriptscriptstyle \pom}}

\newcommand {\apom} {\alpha_{\pomsub}}
                                  
\begin{document}
\title{Energy dependence of exclusive vector-meson production in $ep$
interactions at HERA
}
\author{Aharon Levy
\thanks{Supported in part by the Israel Science Foundation(ISF), by
the German Israel Foundation (GIF) and be the Israel Ministry of Science.}
\address{School of Physics and Astronomy\\ Raymond and Beverly Sackler
Faculty of Exact Science\\ Tel Aviv University, Tel Aviv, Israel}
\address{On behalf on the ZEUS Collaboration}
}
\maketitle
\begin{abstract}
The energy dependence of exclusive vector meson (V) production is
studied using the ZEUS data.  The SU(4) universality of V cross
sections in the $Q^2+M_V^2$ scale is tested.  The energy dependence of the
ratio of the cross sections for V production to the total $\gamma^*p$
cross section is compared with expectations based on pQCD and Regge
approaches.
\end{abstract}
\section{Introduction}
\label{sec-int}

The energy dependence of the cross section for a given reaction is
sensitive to the dominant production mechanism leading to this
particular process.  
In this talk, the energy dependence of the cross section for
exclusive electroproduction of vector-mesons (V) in $ep$ interactions
at HERA is discussed. First, we present an update of an earlier
study~\cite{uni-buda} of the universality of V cross sections, scaled
by the SU(4) factors, when plotted as a function of $Q^2 +
M_V^2$~\cite{uni-vm}, where $Q^2$ is the virtuality of the photon in
the reaction $\gamma^* p \to V p$, and $M_V$ is the mass of 
$V$. The main new ingredient consists of the precise
measurements of the $J/\psi$ cross
section~\cite{psi-gp,psi-dis}. Secondly, we investigate the ratios of
the V production cross sections, $\sigma_V$, to the total $\gamma^\star
p$ cross section, $\sigma_{tot}(\gamma^* p)$, as obtained from the
measurements of the $F_2$ structure function of the proton. The ratios
are compared to expectations based on perturbative QCD (pQCD) and
Regge approaches.
 
In order to minimize normalization uncertainties, we use for this
study only data from the ZEUS
Collaboration~\cite{psi-gp,psi-dis,rho-gp,rho-dis}.

\section{Universality of the $Q^2 + M_V^2$ scale}

It was shown~\cite{uni-buda} that if vector meson cross sections
are weighted by the appropriate SU(4) factor and plotted as function
of $W$ for fixed values of the variable $Q^2+M_V^2$, all light Vs
($\rho^0, \omega, \phi$) lie on a universal curve. However this was
not the case for the $J/\psi$. Since then, the preliminary ZEUS
$J/\psi$ photoproduction cross sections  have been published~\cite{psi-gp}
and new preliminary data on exclusive electroproduction of $J/\psi$,
covering a wider $W$ range, became available~\cite{psi-dis}.

\begin{figure}[h]
\vspace{-0.3cm}
\epsfig{file=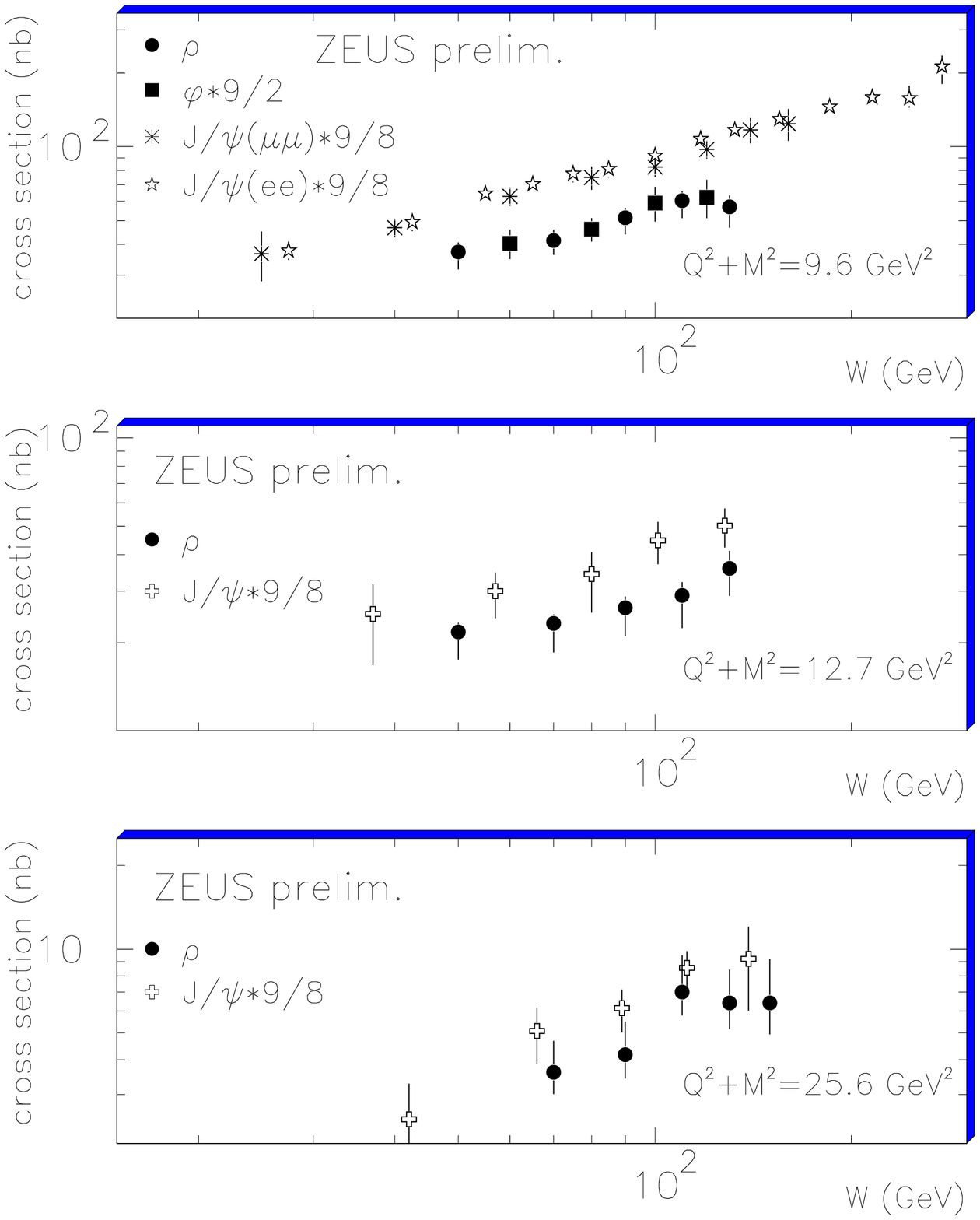,width=0.55\textwidth}
\epsfig{file=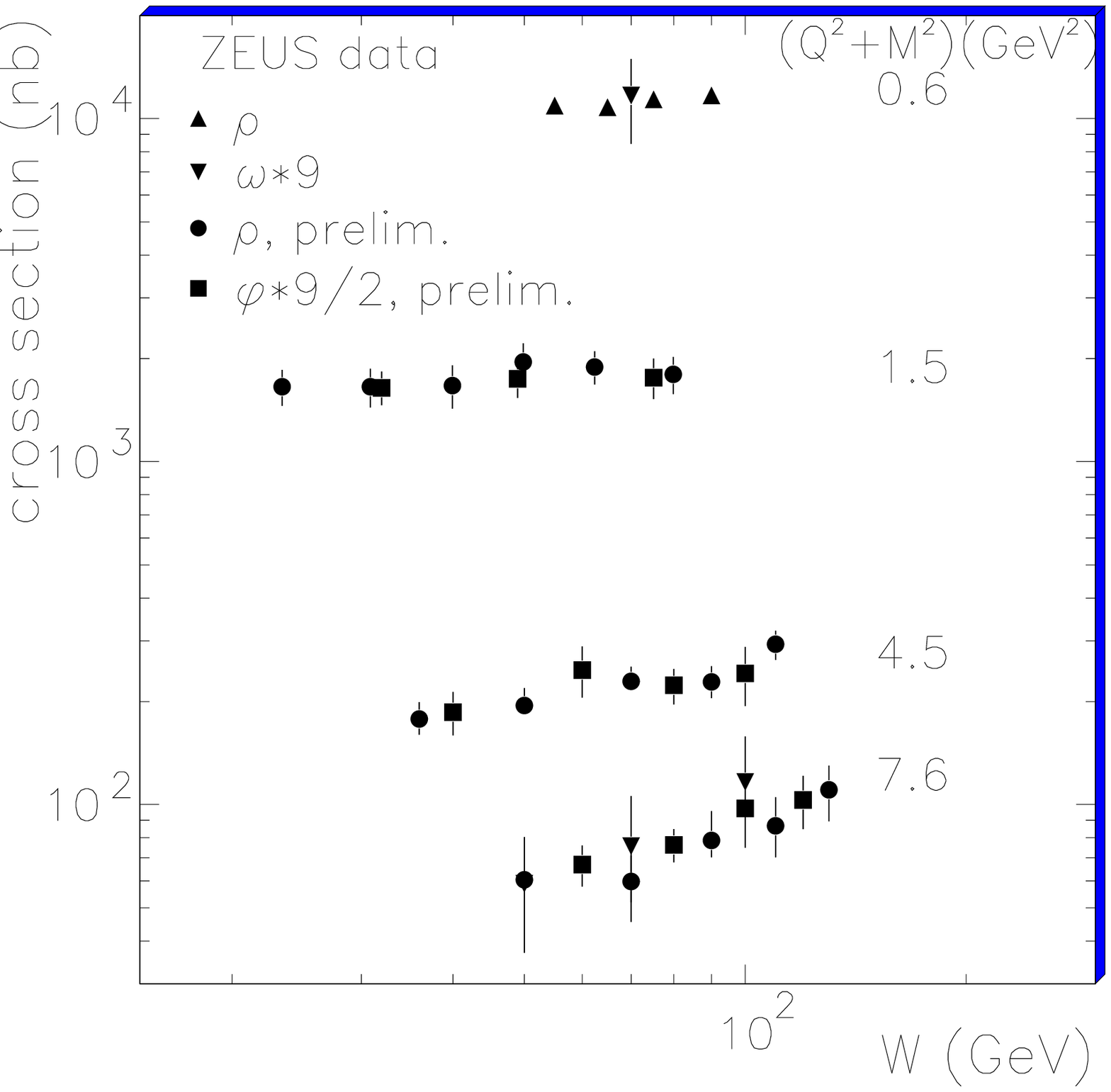,width=0.55\textwidth}
\caption{Comparison of vector meson cross section values, weighted by
the SU(4) factors, plotted as a function of $W$ at fixed
$Q^2+M_V^2$, as indicated in the figure. 
}
\label{fig:su4}
\end{figure}

Fig.~\ref{fig:su4} shows the updated cross sections weighted by the
appropriate SU(4) factors, as function of $W$.
For $Q^2 + M_V^2$ = 9.6 GeV$^2$, the $J/\psi$ cross sections
are about 40\% higher than that of the light Vs. Also at higher scales,
where errors are larger,
the $J/\psi$ cross section values lie systematically above those for
the $\rho^0$. Thus, the updated data confirm the earlier conclusion that
there is no simple universality for V production if $Q^2 +
M_V^2$ is used as a scale.

\section{The ratio $\sigma_V/\sigma_{tot}$}
\label{sec:rv}

The total $\gamma^* p$ cross section, $\sigma_{tot}(\gamma^* p)$,
exhibits a rise with $W$, which becomes steeper with increasing
$Q^2$. It is of interest to compare the $W$ dependence of exclusive V
production to that of the inclusive cross section. To this end we
study the ratio 
\begin{equation}
r_V \equiv \frac{\sigma(\gamma^* p \to V p)}{\sigma_{tot}(\gamma^* p)},
\label{eq:ratio}
\end{equation}
for each of the Vs, as a function of $W$ for fixed $Q^2$. Before
presenting the data we discuss the expectations for $r_V$ using pQCD
and Regge arguments.

\subsection{Expectations for $r_V$ in pQCD}

In pQCD, the forward cross section for longitudinally polarized photons
is expected~\cite{brodsky} to behave as
\begin{equation}
\frac{d\sigma_L}{dt}|_{t=0} \propto \frac{1}{Q^6} \alpha_S^2(Q^2)
[xg(x,Q^2)]^2,
\label{eq:brodsky}
\end{equation}
where $x$ is the Bjorken scaling variable and $xg(x,Q^2)$ is the gluon
density in the proton.

Assuming an exponential $t$ behavior of the form $d\sigma_V/dt \sim
e^{bt}$, one gets the following expectation
\begin{equation}
r_V \propto (1 + \frac{1}{R}) \frac{W^{2\lambda}}{b} \approx 
\frac{W^{2\lambda}}{b}. 
\label{eq:rvqcd}
\end{equation}
In expression~(\ref{eq:rvqcd}) we have used the fact that both the
gluon density distribution and the proton structure function have a
similar $x^{-\lambda(Q^)}$ dependence and that the ratio $R$ of the V
production cross section induced by longitudinal and transverse
virtual photons, $R = \sigma_L/\sigma_T$, increases with $Q^2$ but is
$W$ independent~\cite{rho-dis} and thus can be neglected.

\subsection{Expectations for $r_V$ in Regge phenomenology}

Using Regge phenomenology arguments~\cite{regge}, one expects 
$\sigma_V \sim W^{4(\apom(0)-1)}/b$
and $\sigma_{tot} \sim
W^{2(\apom(0)-1)}$, where $\apom(0)$ is the intercept of the
Pomeron trajectory, and
\begin{equation}
r_V \propto \frac{W^{2(\apom(0)-1)}}{b},
\label{eq:rvregge}
\end{equation} 
which is the same as~(\ref{eq:rvqcd}),
if we write $\lambda = 2(\apom(0)-1)$.

\subsection{Comparison of the measured $r_V$ with expectations}

Both in pQCD and Regge approaches the ratio $r_V$ rises with $W$. The
$W$ dependence is not strongly affected by $b$ since both for the
exclusive electroproduction of $\rho^0$ and the photoproduction of
$J/\psi$ shrinkage was found to be small~\cite{rho-dis,psi-gp}.

When calculating $r_V$ one has to ensure that both cross sections are
taken at the same hard scale. Since for $\sigma_V$ the scale is
usually chosen as $(Q^2 + M_V^2)/4$, $\sigma_{tot}$ is calculated at
that scale.

\begin{figure}[h]
\vspace{-0.3cm}
\epsfig{file=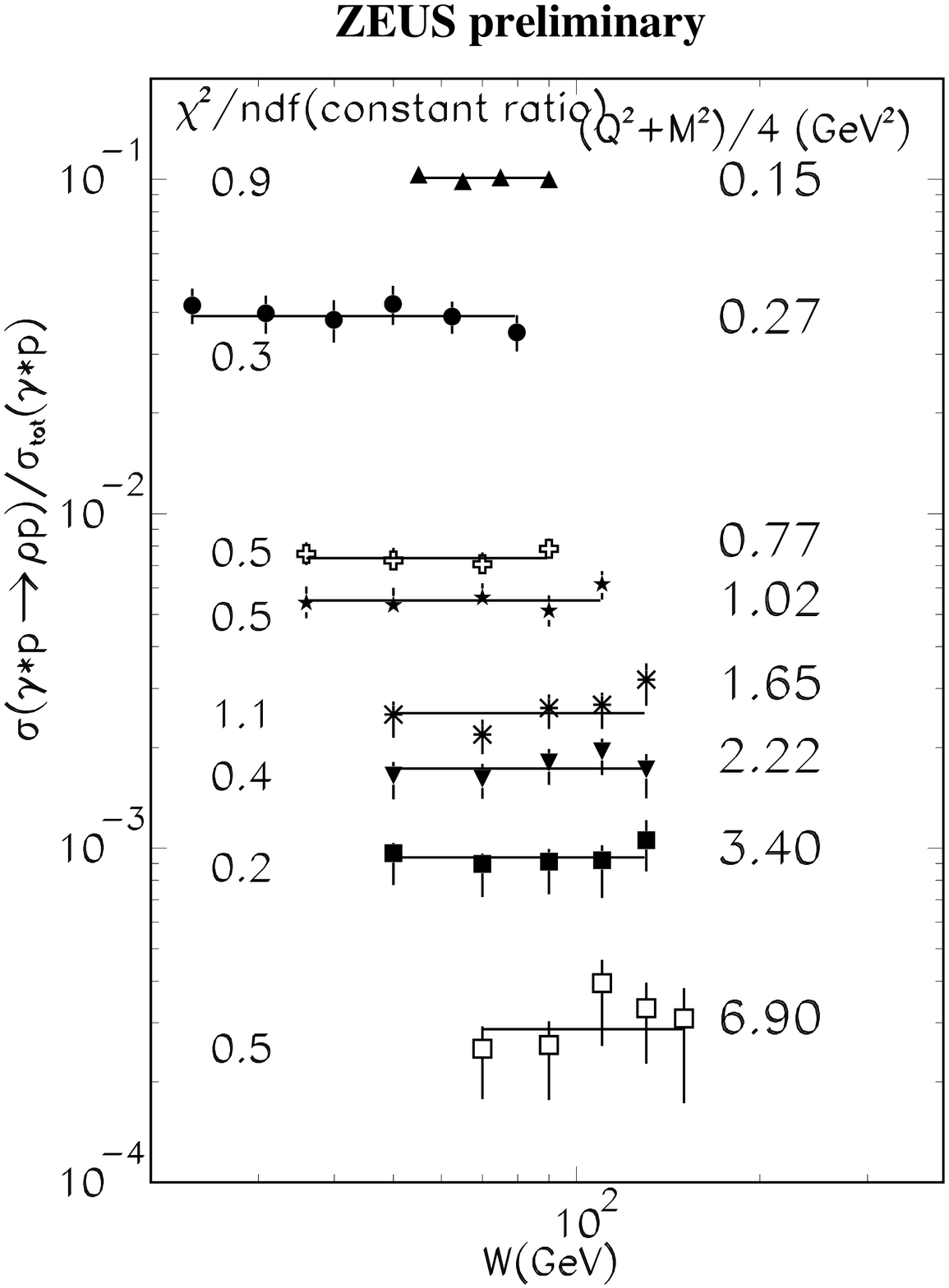,width=0.55\textwidth}
\epsfig{file=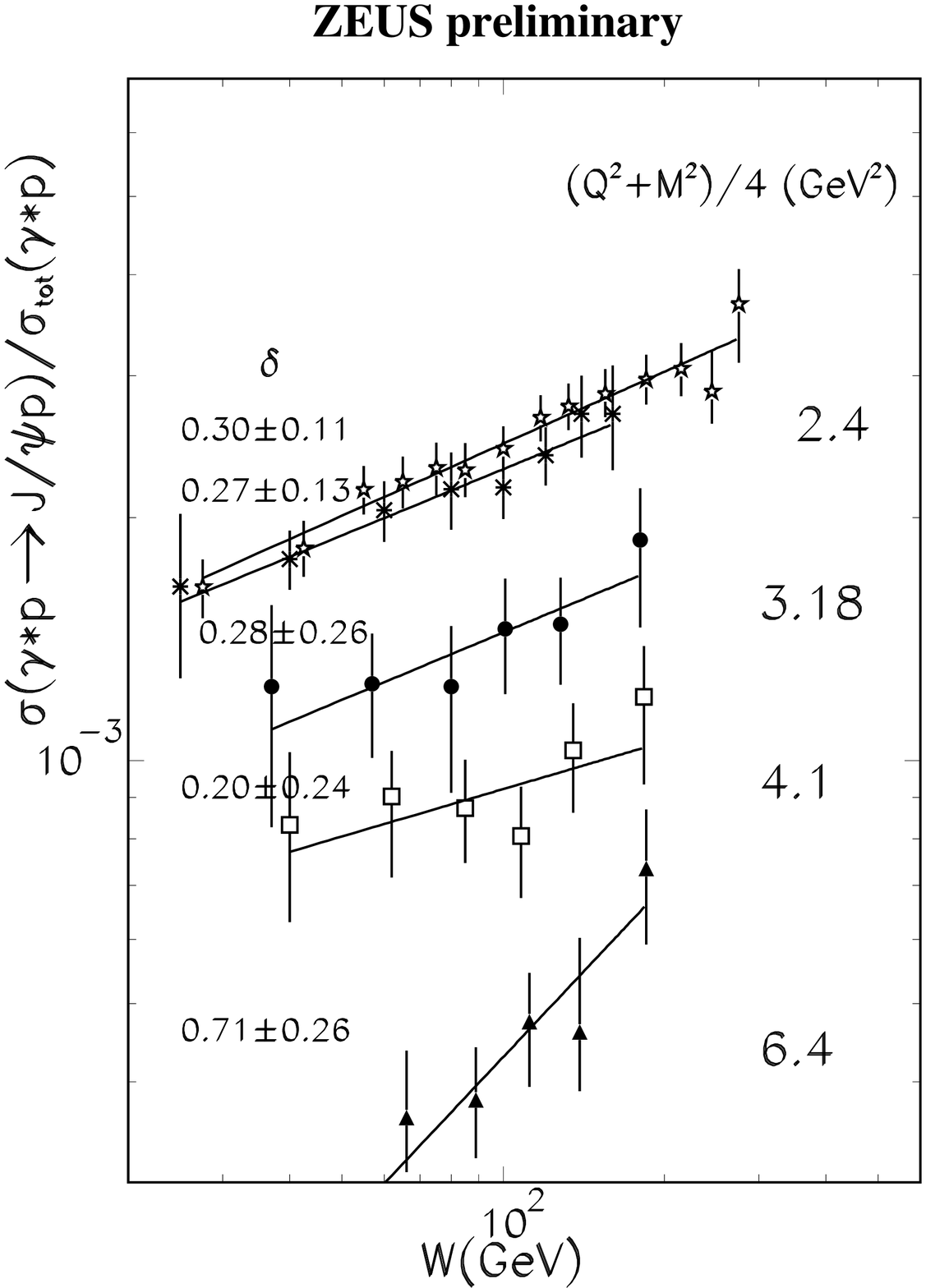,width=0.55\textwidth}
\caption{The ratio $\sigma_V/\sigma_{tot}$ as function of $W$ for 
$V=\rho^0$ (left hand side) and for $V=J/\psi$ (right hand side), at
fixed values of scales, as indicated in the figure.}
\label{fig:rvdata}
\end{figure}

The $W$ dependence of $r_V$ is shown in Fig.~\ref{fig:rvdata} for the
electroproduction of $\rho^0$ and $J/\psi$, for fixed values of the
scale, as indicated in the figure.
While for the $J/\psi$ case the ratio $r_{J/\psi}$ seems to show the
expected rise with $W$, the ratio $r_\rho$ is independent of $W$,
reminiscent of the behavior found in inclusive diffraction at large
$Q^2$~\cite{rdiff}. The behavior of $r_\rho$ cannot be explained by
either pQCD nor by the Regge calculations.

\section{The ratio $\sigma_{tot}^2/\sigma_V$}

Following the discussion in Section~\ref{sec:rv}, it is of interest to
study the ratio $\sigma_{tot}^2/\sigma_V$, as it contains information
about the $Q^2$ behavior in the pQCD approach as well as about the
slope $b$, in both approaches. 

\subsection{$\sigma_{tot}^2/\sigma_V$ in pQCD}

In order to calculate the pQCD expectations for $\sigma_{tot}^2/\sigma_V$, we
include also the terms having a $Q^2$ dependence and obtain
\begin{equation}
\frac{\sigma_{tot}^2}{\sigma_V} \propto \frac{Q^2}{\alpha_S^2(Q^2)(1 +
\frac{1}{R})}b \approx \frac{Q^2}{\alpha_S^2(Q^2)}b.
\label{eq:r2qcd}
\end{equation}
This comes about since $\sigma_{tot} \sim 1/Q^2$ and $\sigma_V \sim
1/Q^6$.  The contribution of $1/R$ is negligible.  Thus, if one
neglects the $Q^2$ variation of $\alpha_S$ in the $Q^2$ range
discussed in this paper,
\begin{equation}
 \frac{\sigma_{tot}^2}{Q^2 \sigma_V} \propto b  
\label{eq:bqcd}
\end{equation}
and, apart from some numerical $W$-independent factor, one obtains the
slope $b$.

\begin{figure}[h]
\vspace{-0.3cm}
\epsfig{file=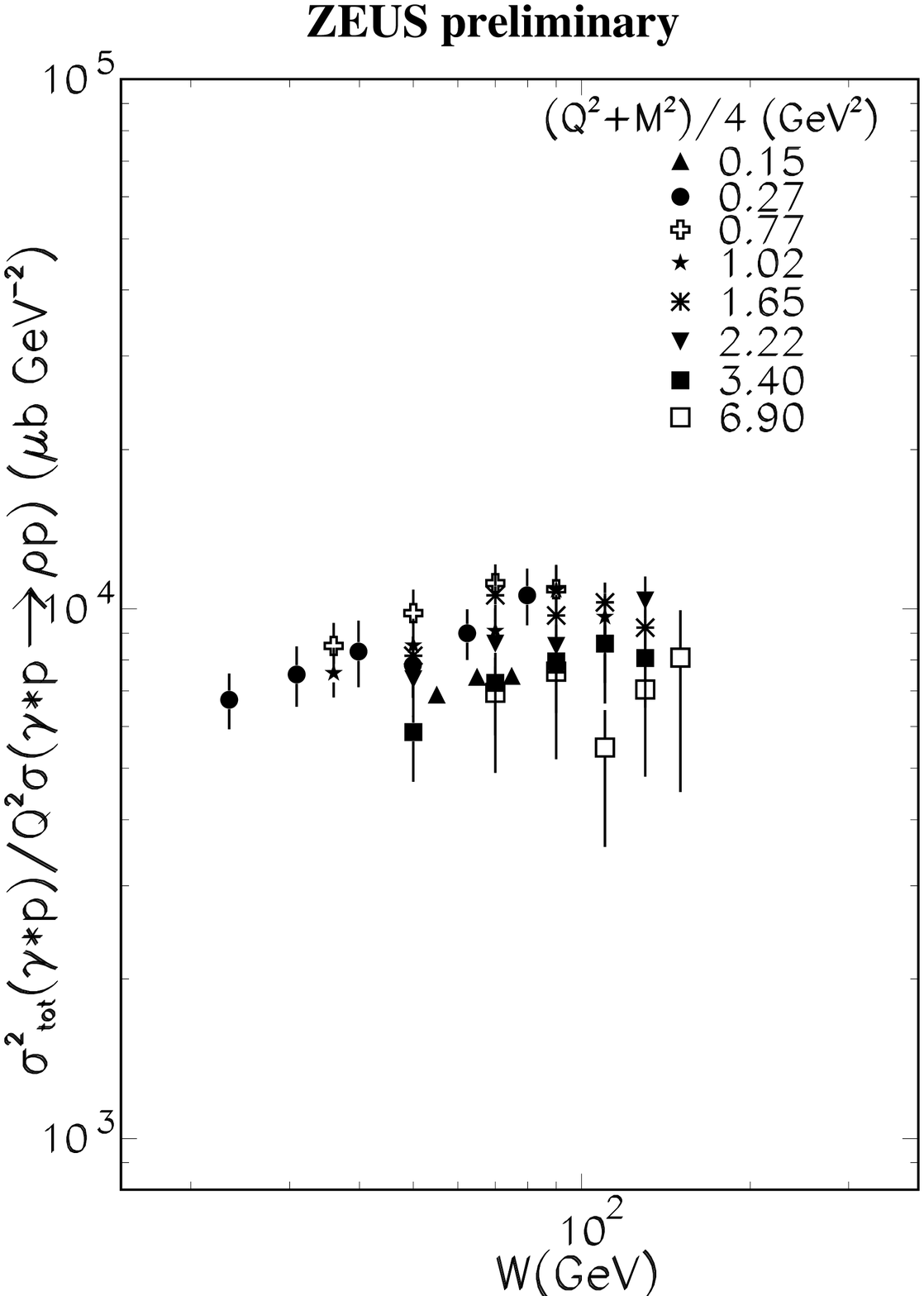,width=0.5\textwidth}
\epsfig{file=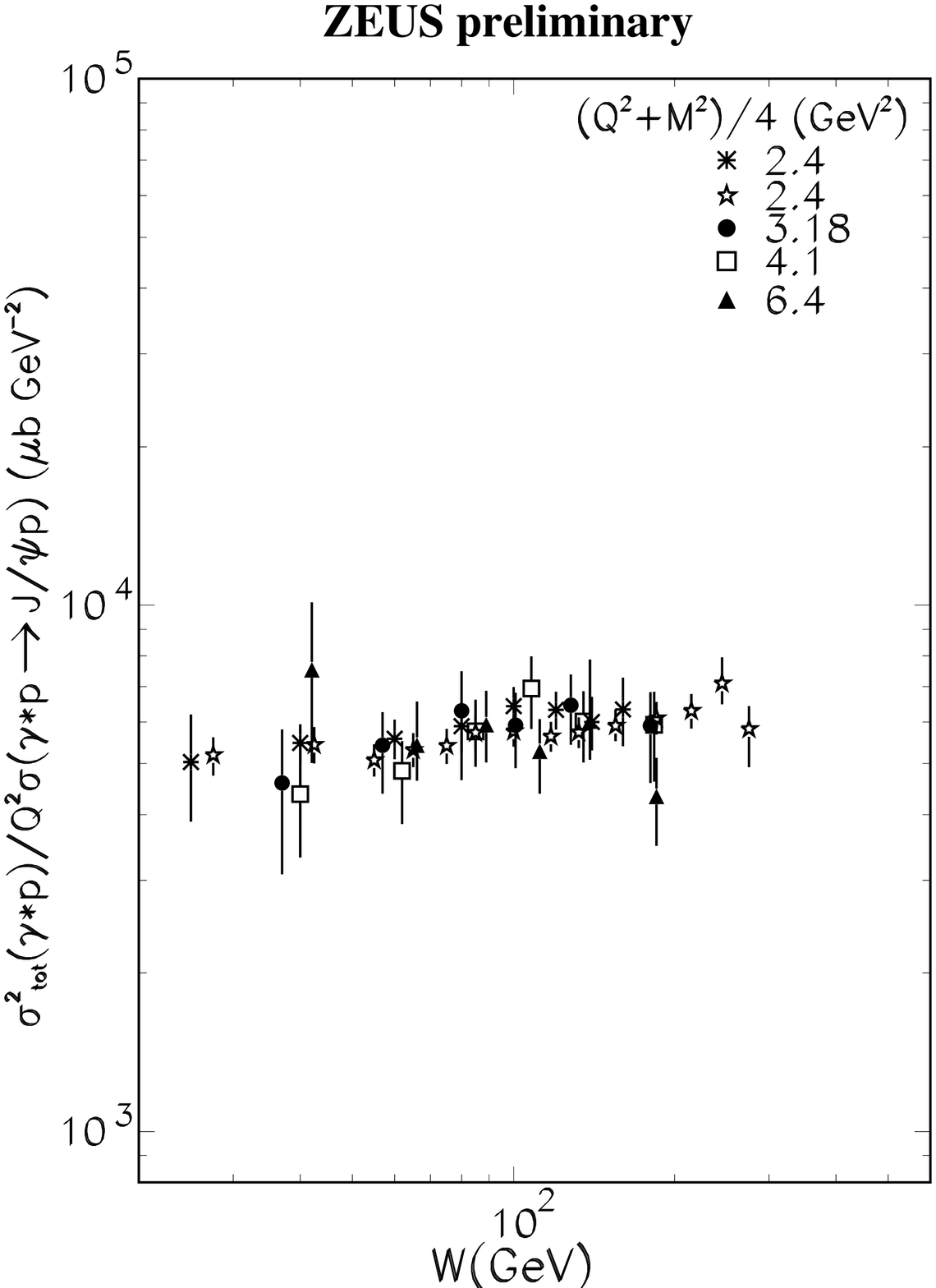,width=0.5\textwidth}
\caption{The ratio $\sigma_{tot}^2/(Q^2 \sigma_V)$ as function of $W$
for $V=\rho^0$ (left hand side) and for $V=J/\psi$ (right hand side),
at fixed values of scales, as indicated in the figure.  }
\label{fig:bqcd}
\end{figure}

Fig.~\ref{fig:bqcd} shows the ratio~(\ref{eq:bqcd}) for the $\rho^0$
and for the $J/\psi$ mesons. 
While in the case of $\rho^0$ one sees a slight $Q^2$ dependence of
the ratio, the ratio is $Q^2$ independent for the $J/\psi$. This
result is in agreement with what is known from direct measurements of
the $Q^2$ dependence of $b$ for both Vs~\cite{rho-fact,psi-dis}.

Reversing the argument, especially in the $J/\psi$ case, where the
lack of $Q^2$ dependence of $b$ is known, Fig.~\ref{fig:bqcd} reflects
the behavior of the propagator term, $1/Q^6$, in $\sigma_V$, as
expected in pQCD or pQCD-inspired models.

\subsection{$\sigma_{tot}^2/\sigma_V$ in Regge phenomenology}

One can use the optical theorem, together with the Vector Dominance
Model (VDM)~\cite{vdm}, to obtain the following relation:
\begin{equation}
\sigma_{tot}^2 = \frac{4\pi \alpha_{em}
\sigma_\rho b}{\frac{4\pi}{\gamma^2_{\gamma\rho}}},
\label{eq:opt}
\end{equation}
where $4\pi/\gamma^2_{\gamma\rho}$ is the strength of the direct
$\gamma$-$\rho^0$ coupling. This relation assumes 
$\rho$-dominance, which is a good assumption as the $\rho^0$
contributes $\approx$ 85\% to the VDM relation. 
If $b$ is measured in units of GeV$^{-2}$,
and a value of 0.5~\cite{bauer} is used for the $\gamma$-$\rho^0$
coupling, one obtains from eq.~(\ref{eq:opt})
\begin{equation}
0.014 \frac{\sigma_{tot}^2}{\sigma_\rho} = b.
\label{eq:bvdm}
\end{equation}

\begin{figure}[h]
\vspace{-0.3cm}
\begin{center}
\epsfig{file=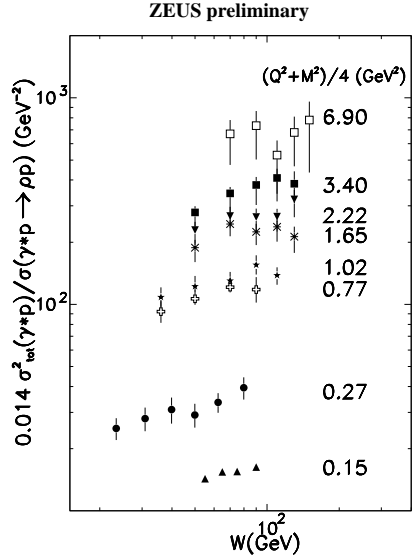,width=0.48\textwidth}
\end{center}
\caption{The ratio $0.014\sigma_{tot}^2/\sigma_\rho$ as function of $W$ 
at fixed values of scales, as indicated in the figure.}
\label{fig:bvdm}
\end{figure}

Fig.~\ref{fig:bvdm} shows the left hand side of
eq.~(\ref{eq:bvdm}), as function of $W$, for fixed values of the scale.
For the $\rho^0$ photoproduction data,
after correcting for the assumed $\rho$-dominance, one
gets good agreement with the value measured~\cite{rho-gp}. 

While the $W$ dependence is consistent with the slight shrinkage
measured~\cite{rho-dis}, the resulting values at higher scales come
out much too high and increase with the scale, contrary to direct
measurements~\cite{rho-fact}.  This is not surprising since in
eq.~(\ref{eq:bvdm}) we have used the value of the direct
$\gamma$-$\rho^0$ coupling at $Q^2$=0. In fact, the direct coupling is
$Q^2$ dependent, and can be expressed as
\begin{equation}
\frac{4\pi}{\gamma^2_{\gamma\rho}(Q^2)} =
\frac{4\pi}{\gamma^2_{\gamma\rho}(0)} f_{\gamma\rho}(Q^2),
\end{equation}
where $f_{\gamma\rho}(Q^2)$ is the $\gamma$-$\rho^0$ form-factor. One
can use the measured values of $b$~\cite{rho-fact} to calculate the
form-factor using eq.~(\ref{eq:opt}),
\begin{equation}
\frac{4\pi}{\gamma^2_{\gamma\rho}(Q^2)} = \frac{4\pi \alpha_{em}
\sigma_\rho}{\sigma_{tot}^2} b(Q^2).
\end{equation}

\begin{figure}[h]
\vspace{-0.3cm}
\begin{center}
\epsfig{file=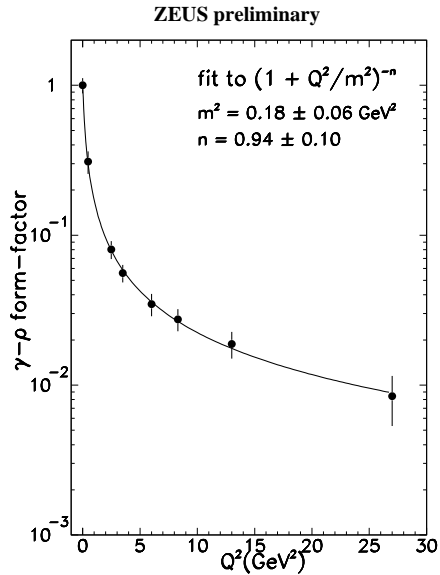,width=0.5\textwidth}
\end{center}
\caption{The $\gamma$-$\rho^0$ form-factor as function of $Q^2$. The
curve is a fit of the form $(1 + Q^2/m^2)^{-n}$ to the data.}
\label{fig:ff}
\end{figure}

Fig.~\ref{fig:ff} shows the resulting values of the $\gamma$-$\rho^0$
form-factor, as function of $Q^2$.
A function of the form $\sim (1 + Q^2/m^2)^{-n}$ was fitted to the
data, resulting in the parameters $m^2 = 0.18 \pm 0.06$ GeV$^2$, and
$n = 0.94 \pm 0.10$. To restore a good agreement with measurements, a
$1/Q^2$ dependence has to be introduced through the form-factor.

\section{Summary}

We can summarize the results of this study as follows:
\begin{itemize}
\item
The SU(4)-weighted cross sections of the light vector mesons lie on a
universal curve when plotted at a scale of $Q^2 + M_V^2$. This is not
the case for the $J/\psi$ meson.
\item
The ratio of the cross section of exclusive $\rho^0$ electroproduction
to that of the total $\gamma^* p$ one is $W$ independent, for fixed
values of $Q^2$. This behavior cannot be explained by either the pQCD
nor by the Regge phenomenology approaches. The ratio rises with $W$
for the $J/\psi$ vector meson, consistent with the expectations.
\item
The ratio $\sigma_{tot}^2/\sigma_V$ shows the $1/Q^6$ dependence
predicted by pQCD for the Vs cross section. This ratio also provides
an indirect way to learn about the $W$ and $Q^2$ behavior of the slope
$b$.
\item
In order to make the VDM approach agree with the data, one would need
to introduce a $\gamma$-$\rho^0$ form factor which has a $1/Q^2$
dependence.
\end{itemize} 

\section*{Acknowledgments}
I would like to thank Andrzej Eskreys, Danuta Kisielewska and the
whole local organizing committee for creating a wonderful atmosphere
in hosting this workshop.

\end{document}